\documentclass[twocolumn,aps]{revtex4}

\usepackage{graphicx} 
\usepackage{dcolumn}  
\usepackage{bm}       

\newcommand{\CNOT}{\textsc{cnot}}

\begin{document}

\title{Scalability of Shor's algorithm with a limited set of rotation gates}

\author{Austin G. Fowler and Lloyd C. L. Hollenberg}

\affiliation{Centre for Quantum Computer Technology,\\ School of
Physics, University of Melbourne, Victoria 3010, AUSTRALIA.}
\date{\today}

\begin{abstract}
Typical circuit implementations of Shor's algorithm involve
controlled rotation gates of magnitude $\pi/2^{2L}$ where $L$ is
the binary length of the integer N to be factored. Such gates
cannot be implemented exactly using existing fault-tolerant
techniques. Approximating a given controlled $\pi/2^{d}$ rotation
gate to within $\delta=O(1/2^{d})$ currently requires both a
number of qubits and number of fault-tolerant gates that grows
polynomially with $d$. In this paper we show that this additional
growth in space and time complexity would severely limit the
applicability of Shor's algorithm to large integers. Consequently,
we study in detail the effect of using only controlled rotation
gates with $d$ less than or equal to some $d_{\rm max}$. It is
found that integers up to length $L_{\rm max} = O(4^{d_{\rm
max}})$ can be factored without significant performance penalty
implying that the cumbersome techniques of fault-tolerant
computation only need to be used to create controlled rotation
gates of magnitude $\pi/64$ if integers thousands of bits long are
desired factored.  Explicit fault-tolerant constructions of such
gates are also discussed.
\end{abstract}


\maketitle

Shor's factoring algorithm \cite{Shor94b,Shor00} is arguably the
driving force of much experimental quantum computing research. It
is therefore crucial to investigate whether the algorithm has a
realistic chance of being used to factor commercially interesting
integers. This paper focuses on the difficulty of implementing the
quantum Fourier transform (QFT) -- an integral part of the
algorithm. Specifically, the controlled $\pi/2^{d}$ rotations that
comprise the QFT are extremely difficult to implement using
fault-tolerant gates protected by quantum error correction (QEC).

To factor an $L$-bit integer $N$, a $2L$-qubit QFT is required
that at first glance involves controlled rotation gates of
magnitude $\pi/2^{2L}$. Prior work on simplifying the QFT so that
it only involves controlled rotation gates of magnitude
$\pi/2^{d_{\rm max}}$ has been performed by Coppersmith
\cite{Copp94} with the conclusion that the length $L_{\rm max}$ of
the maximum length integer that can be factored scales as
$O(2^{d_{\rm max}})$ and that factoring an integer thousands of
bits long would require the implementation of controlled rotations
as small as $\pi/10^{6}$. This paper refines this work with the
conclusion that $L_{\rm max}$ scales as $O(4^{d_{\rm max}})$, with
$\pi/64$ rotations sufficient to enable the factoring of integers
thousands of bits long.

The discussion is organized as follows. In Section~\ref{shor}
Shor's algorithm is revised with emphasis on extracting useful
output from the quantum period finding (QPF) subroutine. This
subroutine is described in detail in this section. In
Section~\ref{aqft} Coppersmith's approximate quantum Fourier
transform (AQFT) is described, followed by
Section~\ref{ftrotation} which outlines the techniques used to
implement the gate set required by the AQFT using only
fault-tolerant gates protected by QEC. In Section~\ref{sVr} the
relationship between the probability of success $s$ of the QPF
subroutine and the period $r$ being sought is investigated.
 In Section~\ref{sVLd}
the relationship between the probability success $s$ and both the
length $L$ of the integer being factored and the minimum angle
controlled rotation $\pi/2^{d_{\rm max}}$ is studied. This is then
used to relate $L_{\rm max}$ to $d_{\rm max}$. Section~\ref{conc}
concludes with a summary of results.

\section{Shor's Algorithm}
\label{shor} Shor's algorithm factors an integer $N=N_{1}N_{2}$ by
finding the period $r$ of a function $f(k)=m^{k}\bmod N$ where
$1<m<N$ and ${\rm gcd}(m,N)=1$. Provided $r$ is even and
$f(r/2)\neq N-1$ the factors are $N_{1}={\rm gcd}(f(r/2)+1,N)$ and
$N_{2}={\rm gcd}(f(r/2)-1,N)$, where ${\rm gcd}$ denotes the
greatest common divisor.  The probability of finding a suitable
$r$ given a randomly selected $m$ such that ${\rm gcd}(m,N)=1$ is
greater than $0.75$ \cite{Niel00}. Thus on average very few values
of $m$ need to be tested to factor $N$.

\begin{figure}
\begin{center}
\resizebox{70mm}{!}{\includegraphics{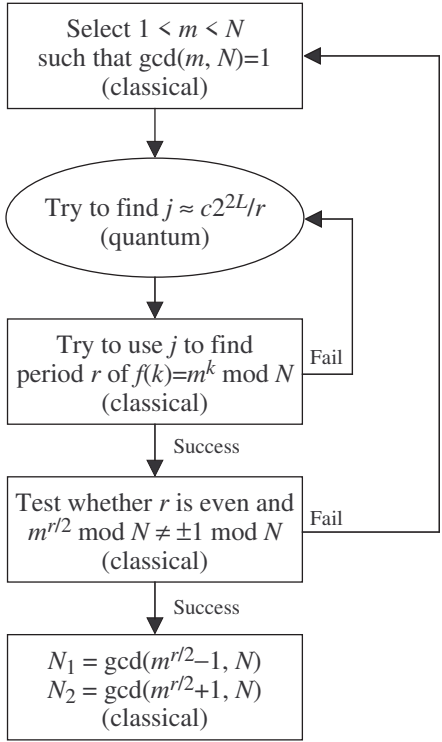}}
\end{center}
\caption{The complete Shor's algorithm including classical pre-
and postprocessing.  The first branch is highly likely to fail,
resulting in many repetitions of the quantum heart of the
algorithm, whereas the second branch is highly likely to succeed.}
\label{figure:flowchart}
\end{figure}

The quantum heart of Shor's algorithm can be viewed as a
subroutine that generates numbers of the form $j \simeq c
2^{2L}/r$. To distinguish this from the necessary classical
pre-and postprocessing, this subroutine will be referred to as QPF
(quantum period finding). For physical reasons, the probability
$s$ that QPF will successfully generate useful data may be quite
low with many repetitions required to work out the period $r$ of a
given $f(k)=m^{k}\bmod N$. Using this terminology, Shor's
algorithm consists of classical preprocessing, potentially many
repetitions of QPF with classical postprocessing and possibly a
small number of repetitions of this entire cycle. This cycle is
summarized in Fig~\ref{figure:flowchart}.

A number of different quantum circuits implementing QPF have been
designed \cite{Vedr96,Goss98,Beau03,Zalk98}.
Table~\ref{table:circuit_comparison} summarizes the number of
qubits required and the depth of each of these circuits. The depth
of a circuit has been defined to be the minimum number of 2-qubit
gates that must be applied sequentially to complete the circuit.
It has been assumed that multiple disjoint 2-qubit gates can be
implemented in parallel, hence the total number of 2-qubit gates
can be significantly greater that the depth.  For example, the
Beauregard circuit has a 2-qubit gate count of $8L^{4}$ to first
order in $L$.
\begin{table}
\begin{tabular}{c|c|c}
Circuit & Qubits & Depth \\
\hline
Beauregard \cite{Beau03} & $2L$ & $32L^{3}$ \\
Vedral \cite{Vedr96} & $5L$ & $240L^{3}$ \\
Zalka \cite{Zalk98}& $\sim 50L$ & $\sim 2^{17}L^{2}$ \\
Gossett \cite{Goss98} & $O(L^{2})$ & $O(L \log L)$ \\
\end{tabular}
\caption{Number of qubits required and circuit depth of different
implementations of Shor's algorithm.  Where possible, figures are
accurate to first order in $L$.}
\label{table:circuit_comparison}
\end{table}
Note that in general the depth of the circuit can be reduced at
the cost of additional qubits.

The underlying algorithm common to each circuit begins by
initializing the quantum computer to a single pure state
$|0\rangle_{2L}|0\rangle_{L}$. Note that for clarity the computer
state has been broken into a $2L$-qubit $k$-register and an
$L$-qubit $f$-register. The meaning of this will become clearer
below.

Step two is to Hadamard transform each qubit in the $k$-register
yielding
\begin{equation}
\label{two}
\frac{1}{2^{L}}\sum_{k=0}^{2^{2L}-1}|k\rangle_{2L}|0\rangle_{L}.
\end{equation}

Step three is to calculate and store the corresponding values of
$f(k)$ in the $f$-register
\begin{equation}
\label{three}
\frac{1}{2^{L}}\sum_{k=0}^{2^{2L}-1}|k\rangle_{2L}|f(k)\rangle_{L}.
\end{equation}
Note that this step requires additional ancilla qubits. The exact
number depends heavily on the circuit used.

Step four can actually be omitted but it explicitly shows the
origin of the period $r$ being sought. Measuring the $f$-register
yields
\begin{equation}
\label{four} \frac{\sqrt{r}}{2^{L}}\sum_{n=0}^{
2^{2L}/r-1}|k_{0}+nr\rangle_{2L}|f_{M}\rangle_{L}
\end{equation}
where $k_{0}$ is the smallest value of $k$ such that $f(k)$ equals
the measured value $f_{M}$.

Step five is to apply the quantum Fourier transform
\begin{equation}
\label{qft1} |k\rangle \rightarrow
\frac{1}{2^{L}}\sum_{j=0}^{2^{2L}-1}\exp(\frac{2\pi i}{2^{2L}} jk)|j\rangle
\end{equation}
to the $k$-register resulting in
\begin{equation}
\label{fivesum}
\frac{\sqrt{r}}{2^{2L}}\sum_{j=0}^{2^{2L}-1}\sum_{p=0}^{
2^{2L}/r-1}\exp(\frac{2\pi i}{2^{2L}}
j(k_{0}+pr))|j\rangle_{2L}|f_{M}\rangle_{L}.
\end{equation}
The probability of measuring a given value of $j$ is thus
\begin{equation}
\label{prj} {\rm Pr}(j,r,L)=\left|\frac{\sqrt{r}}{2^{2L}}\sum_{p=0}^{
2^{2L}/r-1}\exp(\frac{2\pi i}{2^{2L}} jpr)\right|^{2}.
\end{equation}

\begin{figure}
\begin{center}
\resizebox{70mm}{50mm}{\includegraphics{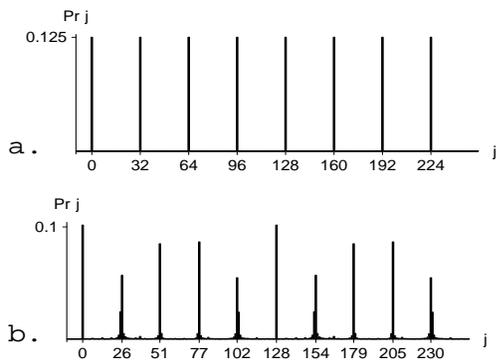}}
\end{center}
\caption{Probability of different measurements $j$ at the end of
quantum period finding with total number of states $2^{2L}=256$
and a.) period $r=8$, b.) period $r=10$.} \label{figure:period}
\end{figure}

If $r$ divides $2^{2L}$ Eq~(\ref{prj}) can be evaluated exactly.
In this case the probability of observing $j=c2^{2L}/r$ for some
integer $0\leq c<r$ is $1/r$ whereas if $j\neq c2^{2L}/r$ the
probability is 0. This situation is illustrated in
Fig~\ref{figure:period}(a). However if $r$ divides $2^{2L}$ exactly
a quantum computer is not needed as $r$ would then be a power of 2
and easily calculable. When $r$ is not a power of 2 the perfect
peaks of Fig~\ref{figure:period}(a) become slightly broader as
shown in Fig~\ref{figure:period}(b). All one can then say is that
with high probability the value $j$ measured will satisfy $j\simeq
c2^{2L}/r$ for some $0\leq c<r$.

Given a measurement $j\simeq c2^{2L}/r$ with $c\neq 0$, classical
postprocessing is required to extract information about $r$. The
process begins with a continued fraction expansion. To illustrate,
consider factoring 143 ($L=8$). Suppose we choose $m$ equal 2 and
the output $j$ of QPF is 31674. The relation $j\simeq c2^{2L}/r$
becomes $31674\simeq c65536/r$. The continued fraction expansion
of $c/r$ is
\begin{equation}
\label{contfrac854}
\frac{31674}{65536}=\frac{1}{\frac{32768}{15837}}=\frac{1}{2+\frac{1094}{15837}}=\frac{1}{2+\frac{1}{14+\frac{1}{2+\frac{1}{10+1/52}}}}.
\end{equation}
The continued fraction expansion of any number between 0 and 1 is
completely specified by the list of denominators which in this
case is $\{2,14,2,10,52\}$.  The $n$th convergent of a continued
fraction expansion is the proper fraction equivalent to the first
$n$ elements of this list.  An introductory exposition and further
properties of continued fractions are described in
Ref~\cite{Niel00}.
\begin{eqnarray}
\label{convergeants427}
\{2\} & = & \frac{1}{2} \nonumber \\
\{2,14\} & = & \frac{14}{29} \nonumber \\
\{2,14,2\} & = & \frac{29}{60} \nonumber \\
\{2,14,2,10\} & = & \frac{304}{629} \nonumber \\
\{2,14,2,10,52\} & = & \frac{15837}{32768}
\end{eqnarray}
The period $r$ can be sought by substituting each denominator into
the function $f(k)= 2^{k} \bmod 143$.
With high probability only the largest denominator less than
$2^{L}$ will be of interest.
In this case $2^{60}\bmod 143=1$ and hence $r=60$.

Two modifications to the above are required. Firstly, if $c$ and
$r$ have common factors, none of the denominators will be the
period but rather one will be a divisor of $r$.
After repeating QPF a number of times, let $\{j_{m}\}$ denote the
set of measured values.
Let $\{c_{mn}/d_{mn}\}$ denote the set of convergents associated
with each measured value $\{j_{m}\}$.
If a pair $c_{mn}$, $c_{m'n'}$ exists such that ${\rm gcd}(c_{mn},
c_{m'n'})=1$ and $d_{mn}$, $d_{m'n'}$ are divisors of $r$ then
$r={\rm lcm}(d_{mn}, d_{m'n'})$, where ${\rm lcm}$ denotes the
least common multiple.
It can be shown that given any two divisors $d_{mn}$, $d_{m'n'}$
with corresponding $c_{mn}$, $c_{m'n'}$ the probability that ${\rm
gcd}(c_{mn},c_{m'n'})=1$ is at least $1/4$ \cite{Niel00}.
Thus only $O(1)$ different divisors are required.
In practice, it will not be known which denominators are divisors
so every pair $d_{mn}$, $d_{m'n'}$ with ${\rm gcd}(c_{mn},
c_{m'n'})=1$ must be tested.

The second modification is simply allowing for the output $j$ of
QPF being useless. Let $s$ denote the probability that $j=\lfloor
c2^{2L}/r\rfloor$ or $\lceil c2^{2L}/r\rceil$ for some $0<c<r$
where $\lfloor \rfloor$, $\lceil \rceil$ denote rounding down and
up respectively. Such values of $j$ will be called useful as the
denominators of the associated convergents are guaranteed to
include a divisor of $r$. The period $r$ sought can always be
found provided $O(1/s)$ runs of QFT can be performed.

To summarize, as each new value $j_{m}$ is measured, the
denominators $d_{mn}$ less than $2^{L}$ of the convergents of the
continued fraction expansion of $j_{m}/2^{2L}$ are substituted
into $f(k)=m^{k} \bmod N$ to determine whether any $f(d_{mn})=1$
which would imply that $r=d_{mn}$. If not, every pair $d_{mn}$,
$d_{m'n'}$ with associated numerators $c_{mn}$, $c_{m'n'}$
satisfying  ${\rm gcd}(c_{mn}, c_{m'n'})=1$ must be tested to see
whether $r={\rm lcm}(d_{mn}, d_{m'n'})$. Note that as shown in
Fig~\ref{figure:flowchart}, if $r$ is even or $m^{r/2}\bmod N =
\pm 1\bmod N$ then the entire process needs to be repeated $O(1)$
times. Thus Shor's algorithm always succeeds provided $O(1/s)$
runs of QFT can be performed.

\section{Approximate Quantum Fourier Transform}
\label{aqft}

\begin{figure}
\begin{center}
\resizebox{!}{55mm}{\includegraphics{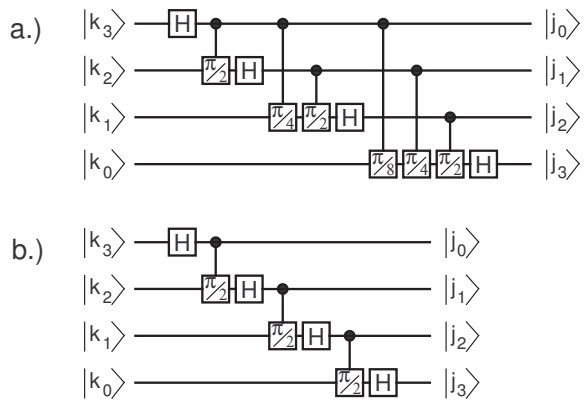}}
\end{center}
\caption{Circuit for a 4-qubit a.) quantum Fourier transform b.)
approximate quantum Fourier transform with $d_{\rm max}=1$}
\label{figure:qftboth}
\end{figure}

A circuit that implements the QFT of Eq~(\ref{qft1}) is shown in
Fig~\ref{figure:qftboth}(a). Note the use of controlled rotations
of magnitude $\pi/2^{d}$. In matrix notation these 2-qubit
operations correspond to
\begin{equation}
\label{contphase}
\left( \begin{array}{cccc}
1 & 0 & 0 & 0 \\
0 & 1 & 0 & 0 \\
0 & 0 & 1 & 0 \\
0 & 0 & 0 & e^{i\pi/2^{d}}
\end{array} \right).
\end{equation}

The approximate QFT (AQFT) circuit is very similar with just the
deletion of rotation gates with $d$ greater than some $d_{\rm
max}$. For example, Fig~\ref{figure:qftboth}(b) shows an AQFT with
$d_{\rm max}=1$. Let $[j]_{m}$ denote the $m$th bit of $j$. The
AQFT equivalent to Eq~(\ref{qft1}) is \cite{Copp94}
\begin{equation}
\label{aqfteq}
|k\rangle \rightarrow
\frac{1}{\sqrt{2^{2L}}}\sum_{j=0}^{2^{2L}-1}|j\rangle
\exp(\frac{2\pi i}{2^{2L}} {\textstyle
\tilde{\sum}_{mn}}[j]_{m}[k]_{n}2^{m+n})
\end{equation}
where $\tilde{\sum}_{mn}$ denotes a sum over all $m$, $n$ such
that $0\leq m,n<2L$ and $2L-d_{\rm max}+1\leq m+n<2L$. It has been
shown by Coppersmith that the AQFT is a good approximation of the
QFT \cite{Copp94} in the sense that the phase of individual
computational basis states in the output of the AQFT differ in
angle from those in the output of the QFT by at most $2\pi
L2^{-d_{\rm max}}$. The purpose of this paper is to investigate in
detail the effect of using the AQFT in Shor's algorithm.

\section{Fault-Tolerant Construction of Small Angle Rotation
Gates}
\label{ftrotation}

\begin{figure}
\begin{center}
\resizebox{70mm}{!}{\includegraphics{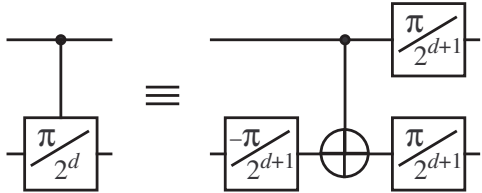}}
\end{center}
\caption{Decomposition of a controlled rotation into single-qubit
gates and a \CNOT.}
\label{figure:phasedec}
\end{figure}

When the 7-qubit Steane code \cite{Stea96,Cald95,Niel00} and its
concatenated generalizations are used to do computation, only the
limited set of gates \CNOT, Hadamard ($H$), $X$, $Z$, $S$ and
$S^{\dag}$ can be implemented easily, where
\begin{equation}
\label{Sgate}
S = \left( \begin{array}{cc}
1 & 0 \\
0 & i \\
\end{array} \right).
\end{equation}
Complicated circuits of depth in the hundreds and requiring a
minimum of 22 qubits are required to implement the $T$ and
$T^{\dag}$ gates \cite{Niel00}
\begin{equation}
\label{Tgate}
T = \left( \begin{array}{cc}
1 & 0 \\
0 & e^{i\pi/4} \\
\end{array} \right).
\end{equation}
Note however that if it is acceptable to add an additional 15
qubits for every $T$ and $T^{\dag}$ gate in a sequence of
fault-tolerant single-qubit gates (see for example Eq
(\ref{U31})), the effective depth of each $T$ and $T^{\dag}$ gate
circuit can be reduced to 2.  Together, the set \CNOT, $H$, $X$,
$Z$, $S$, $S^{\dag}$, $T$ and $T^{\dag}$ enables the
implementation of arbitrary 1- and 2-qubit gates via the
Solovay-Kitaev theorem \cite{Kita97,Niel00}. For example, the
controlled $\pi/2^{d}$ gate can be decomposed into a single \CNOT\
and three single-qubit rotations as shown in
Fig~\ref{figure:phasedec}.
Approximating single-qubit $\pi/2^{d}$ rotations using the
fault-tolerant gate set is much more difficult.  For convenience,
such rotations will henceforth be denoted by $R_{2^{d}}$.  The
simplest (least number of fault-tolerant gates) approximation of
the $R_{128}$ single-qubit rotation gate that is more accurate
than simply the identity matrix is the 31 gate product
\begin{eqnarray}
\label{U31}
U_{31} & = & HTHT^{\dag}HTHTHTHT^{\dag}HT^{\dag}HT\nonumber \\
& & HTHT^{\dag}HT^{\dag}HTHT^{\dag}HT^{\dag}HT^{\dag}H.
\end{eqnarray}
Eq~(\ref{U31}) was determined via an exhaustive search minimizing
the metric
\begin{equation}
\label{dist}
{\rm dist}(U,V) = \sqrt{\frac{2- |{\rm tr}(U^{\dag}V)|}{2}}
\end{equation}
The rationale of Eq~(\ref{dist}) is that if $U$ and $V$ are
similar, $U^{\dag}V$ will be close to the identity matrix
(possibly up to some global phase) and the absolute value of the
trace will be close to 2. By subtracting this absolute value from
2 and dividing by 2 a number between 0 and 1 is obtained.
The overall square root is required to ensure that the triangle
inequality
\begin{equation}
\label{triangle}
{\rm dist}(U,W) \leq {\rm dist}(U,V)+{\rm dist}(V,W)
\end{equation}
is satisfied.

The identity matrix is a good approximation of $R_{128}$ in the
sense that ${\rm dist}(R_{128},I) = 8.7\times 10^{-3}$.
Eq~(\ref{U31}) is only slightly better with ${\rm
dist}(R_{128},U_{31}) = 8.1\times 10^{-3}$. A 46 gate sequence has
been found satisfying ${\rm dist}(R_{128},U_{46}) = 7.5\times
10^{-4}$. Note that this is still only 10 times better than doing
nothing. Further investigation of the properties of fault-tolerant
approximations of arbitrary single-qubit unitaries will be
performed in the near future. For the present discussion it
suffices to know that the number of gates grows somewhere between
linearly and quadratically with $\ln(1/{\delta})$ \cite{Niel00}
where $\delta={\rm dist}(R,U)$, $R$ is the rotation being
approximated, and $U$ is the approximating product of
fault-tolerant gates (the exact scaling is not known). In
particular, this means that approximating a rotation gate
$R_{2^{d}}$ with accuracy $\delta=1/2^{d}$ requires a number of
gates that grows linearly or quadratically with $d$.

In addition to the inconveniently large number of fault-tolerant
gates $n_{\delta}$ required to achieve a given approximation
$\delta$, each individual gate in the approximating sequence must
be implemented with probability of error $p$ less than
$O(\delta/n_{\delta})$. Note that $\delta$ is not a probability of
error but rather a measure of the distance between the ideal gate
and the approximating product so this relationship is not exact.
If the required probability $p\sim
\delta/n_{\delta}=1/(n_{\delta}2^{d})$ is too small to be achieved
using a single level of QEC, the technique of concatenated QEC
must be used. Roughly speaking, if a given gate can be implemented
with probability of error $p$, adding an additional level of
concatenation \cite{Knil96b} leads to an error rate of $cp^{2}$
where $c<1/p$. If the Steane code is used with seven qubits for
the code and an additional five qubits for fault-tolerant
correction, every additional level of concatenation requires 12
times as many qubits. This implies that if a gate is to be
implemented with accuracy $1/(n_{\delta}2^{d})$, the number of
qubits $q$ scales as $O(d^{\ln_{2} 12}) = O(d^{3.58})$. While this
is a polynomial number of qubits, for even moderate values of $d$
this leads to thousands of qubits being used to achieve the
required gate accuracy.

Given the complexity of implementing $T$ and $T^{\dag}$ gates, the
number of fault-tolerant gates required to achieve good
approximations of arbitrary rotation gates and the large number of
qubits required to achieve sufficiently reliable operation, it is
clear that for practical reasons the use of $\pi/2^{d}$ rotations
must be restricted to those with very small $d$.

\section{Dependence of Output Reliability on Period of $f(k)=m^{k}\bmod N$}
\label{sVr}

\begin{figure}[b]
\begin{center}
\resizebox{78mm}{!}{\includegraphics{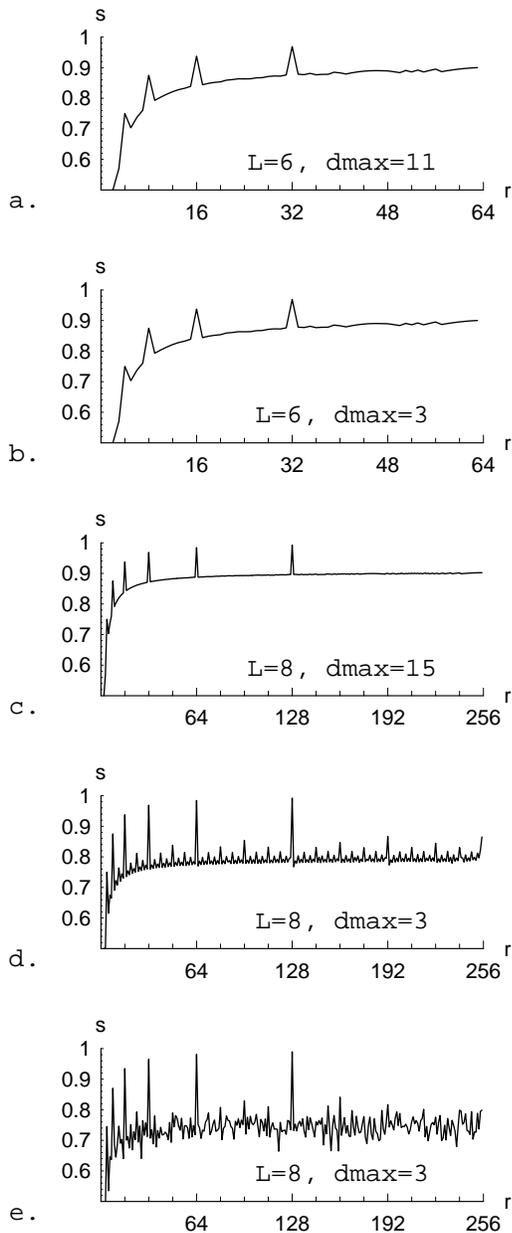}}
\end{center}
\caption{Probability $s$ of obtaining useful output from quantum
period finding as a function of period $r$ for different integer
lengths $L$ and rotation gate restrictions $\pi/2^{d_{\rm max}}$.
The effect of using inaccurate controlled rotation gates
($\sigma=\pi/32$) is shown in (e).} \label{figure:Ldarray}
\end{figure}

Different values of $r$ (the period of $f(k)=x^{k}\bmod N$) imply
different probabilities $s$ that the value $j$ measured at the end
of QPF will be useful (see Fig~\ref{figure:flowchart}). In
particular, as discussed in Section~\ref{shor} if $r$ is a power
of 2 the probability of useful output is much higher (see
Fig~\ref{figure:period}). This section investigates how sensitive
$s$ is to variations in $r$. Recall Eq~(\ref{prj}) for the
probability of measuring a given value of $j$. When the AQFT
(Eq~(\ref{aqfteq}) is used this becomes
\begin{equation}
\label{aqftprj}
\begin{array}{l}
{\rm Pr}(j,r,L,d_{\rm max}) = \\
{\displaystyle \left|\frac{\sqrt{r}}{2^{2L}}\sum_{p=0}^{
2^{2L}/r-1}\exp(\frac{2\pi i}{2^{2L}}{\textstyle
\tilde{\sum}_{mn}[j]_{m}[pr]_{n}2^{m+n}})\right|^{2}}
\end{array}
\end{equation}
The probability $s$ of useful output is thus
\begin{equation}
\label{s}
s(r,L,d_{\rm max})=\sum_{\{{\rm useful}~j\}}{\rm
Pr}(j,r,L,d_{\rm max})
\end{equation}
where $\{{\rm useful}~j\}$ denotes all $j=\lfloor
c2^{2L}/r\rfloor$ or $\lceil c2^{2L}/r\rceil$ such that $0<c<r$.
Fig~\ref{figure:Ldarray} shows $s$ for $r$ ranging from 2 to
$2^{L}-1$ and for various values of $L$ and $d_{\rm max}$. The
decrease in $s$ for small values of $r$ is more a result of the
definition of $\{{\rm useful}~j\}$ than an indication of poor
data. When $r$ is small there are few useful values of $j\simeq
c2^{2L}/r\rceil$, $0<c<r$ and a large range states likely to be
observed around each one resulting superficially in a low
probability of useful output $s$ as $s$ is the sum of the
probabilities of observing only values $j=\lfloor
c2^{2L}/r\rfloor$ or $\lceil c2^{2L}/r\rceil$, $0<c<r$. However,
in practice values much further from $j\simeq c2^{2L}/r$ can
be used to obtain useful output. For example if $r=4$ and
$j=16400$ the correct output value (4) can still be determined
from the continued fraction expansion of $16400/65536$ which is
far from the ideal case of $16384/65536$. To simplify subsequent
analysis each pair $(L, d_{\rm max})$ will from now on be
associated with $s(2^{L-1}+2,L,d_{\rm max})$ which corresponds to
the minimum value of $s$ to the right of the central peak. The
choice of this point as a meaningful characterization of the
entire graph is justified by the discussion above.

For completeness, Fig~\ref{figure:Ldarray}(e) shows the case of
noisy controlled rotation gates of the form
\begin{equation}
\label{contphase_delta} \left( \begin{array}{cccc}
1 & 0 & 0 & 0 \\
0 & 1 & 0 & 0 \\
0 & 0 & 1 & 0 \\
0 & 0 & 0 & e^{i(\pi/2^{d}+\delta)}
\end{array} \right).
\end{equation}
where $\delta$ is a normally distributed random variable of
standard deviation $\sigma$. This has been included to simulate
the effect of using approximate rotation gates built out of a
finite number of fault-tolerant gates.  The general form and
probability of successful output can be seen to be similar despite
$\sigma=\pi/32$. This $\sigma$ corresponds to $\pi/2^{d_{\rm
max}+2}$. For a controlled $\pi/64$ rotation, single-qubit
rotations of angle $\pi/128$ are required, as shown in
Fig~\ref{figure:phasedec}. Fig~\ref{figure:Ldarray}(e) implies
that it is acceptable for this rotation to be implemented within
$\pi/512$, implying
\begin{equation}
\label{contphase_U} U = \left( \begin{array}{cccc}
1 & 0 \\
0 & e^{i(\pi/128+\pi/512)}
\end{array} \right)
\end{equation}
is an acceptable approximation of $R_{128}$.  Given that ${\rm
dist}(R_{128}, U) = 2.1\times 10^{-3}$, the 46 fault-tolerant gate
approximation of $R_{128}$ mentioned above is adequate.

\section{Dependence of Output Reliability on Integer Length and Gate Restrictions}
\label{sVLd}

\begin{figure*}
\begin{tabular}{cc}
\resizebox{65mm}{!}{\includegraphics{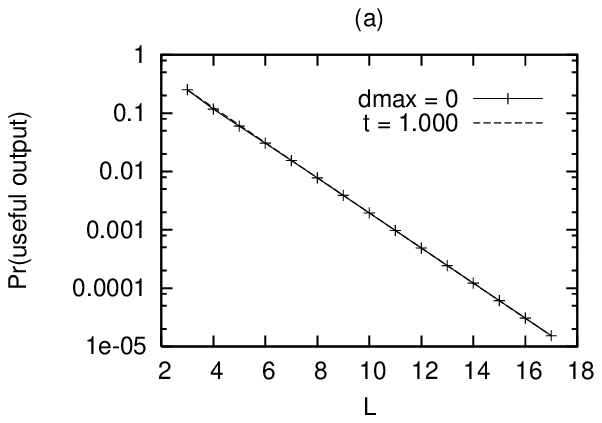}} & \resizebox{65mm}{!}{\includegraphics{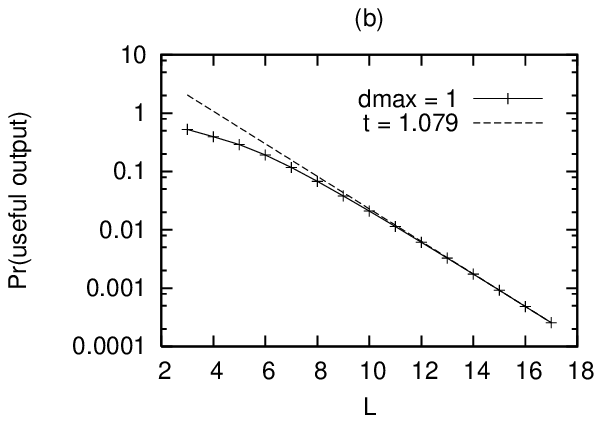}}\\
\resizebox{65mm}{!}{\includegraphics{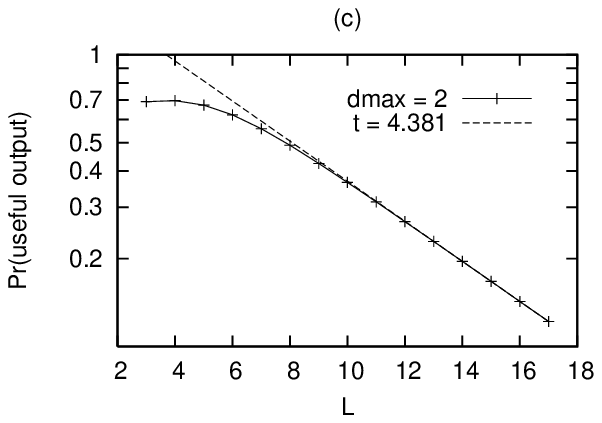}} & \resizebox{65mm}{!}{\includegraphics{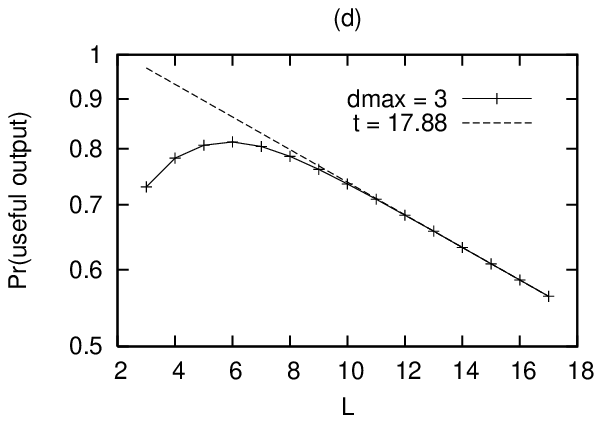}}\\
\resizebox{65mm}{!}{\includegraphics{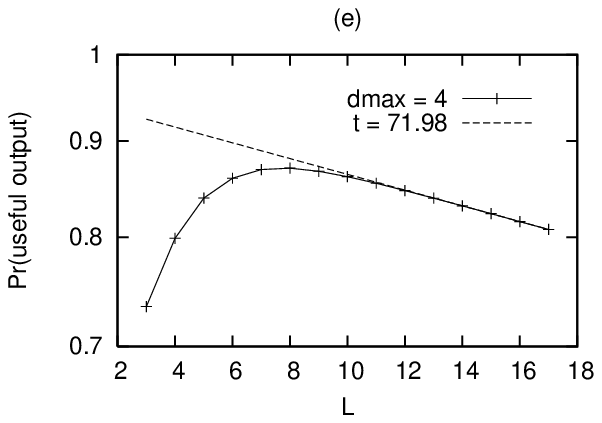}} & \resizebox{65mm}{!}{\includegraphics{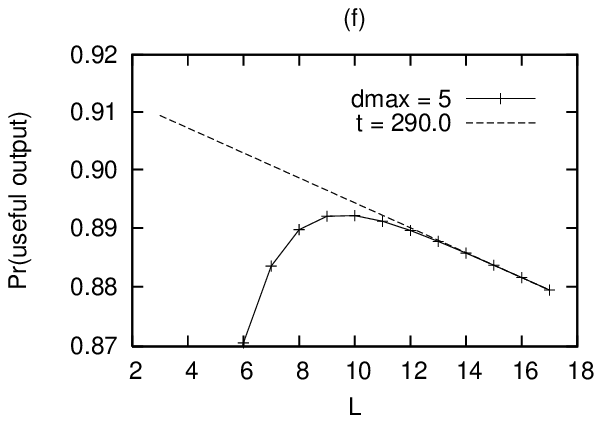}}\\
\resizebox{65mm}{!}{\includegraphics{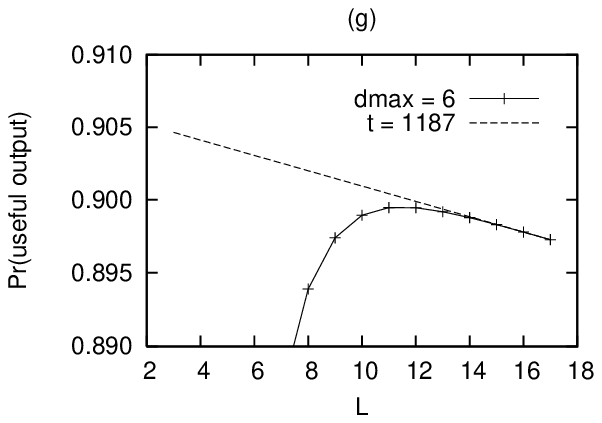}} & \resizebox{65mm}{!}{\includegraphics{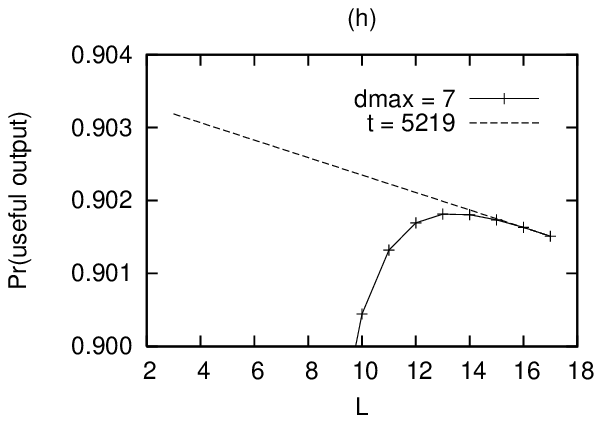}}\\
\end{tabular}
\caption{Dependence of the probability of useful output from the
quantum part of Shor's algorithm on the length $L$ of the integer
being factored for different levels of restriction of controlled
rotation gates of angle $\pi/2^{d_{\rm max}}$}.
\label{figure:darray}
\end{figure*}

In order to determine how the probability of useful output $s$
depends on both the integer length $L$ and the minimum allowed
controlled rotation $\pi/2^{d_{\rm max}}$, Eq~(\ref{s}) was solved
with $r=2^{L-1}+2$ as discussed in Section \ref{sVr}.
Fig~\ref{figure:darray} contains semilog plots of $s$ versus $L$
for different values of $d_{\rm max}$. Note that Eq~(\ref{s})
grows exponentially more difficult to solve as $L$ increases.

For $d_{\rm max}$ from 0 to 5, the exponential decrease of $s$
with increasing $L$ is clear.  Asymptotic lines of best fit of the
form
\begin{equation}
\label{fit}
s \propto 2^{-L/t}
\end{equation}
have been shown. Note that for $d_{\rm max}>0$, the value of $t$
increases by greater than a factor of 4 when $d_{\rm max}$
increases by 1.  This enables one to generalize Eq~(\ref{fit}) to
an asymptotic lower bound valid for all $d_{\rm max}>0$
\begin{equation}
\label{genfit} s \propto 2^{-L/4^{d_{\rm max}-1}}
\end{equation}
with the constant of proportionality approximately equal to 1.

Keeping in mind that the required number of repetitions of QPF is
$O(1/s)$, one can relate $L_{\rm max}$ to $d_{\rm max}$ by
introducing an additional parameter $f_{\rm max}$ characterizing
the acceptable number of repetitions of QPF
\begin{equation}
\label{Leq} L_{\rm max}\simeq 4^{d_{\rm max}-1}\log_{2}f_{\rm max}.
\end{equation}

Available RSA \cite{Rive78} encryption programs such as PGP
typically use integers of length $L$ up to 4096. The circuit in
\cite{Vedr96} runs in $150L^3$ steps when an architecture that can
interact arbitrary pairs of qubits in parallel is assumed and
fault-tolerant gates are used. Note that to first order in $L$ the
number of steps does not increase as additional levels of QEC are
used. Thus $\sim$$10^{13}$ steps are required to perform a single
run of QPF. On an electron spin or charge quantum computer
\cite{Burk00,Holl03} running at 10GHz this corresponds to
$\sim$$15$ minutes of computing. If we assume $\sim$24 hours of
computing is acceptable then $f_{\rm max}\sim 10^2$. Substituting
these values of $L_{\rm max}$ and $f_{\rm max}$ into
Eq~(\ref{Leq}) gives $d_{\rm max}=6$ after rounding up. Thus
provided controlled $\pi/64$ rotations can be implemented
accurately, implying the need to accurately implement $\pi/128$
single-qubit rotations, it is conceivable that a quantum computer
could one day be used to break a 4096-bit RSA encryption in a
single day.

\section{Conclusion}
\label{conc}

We have demonstrated the robustness of Shor's algorithm when a
limited set of rotation gates is used. The length $L_{\rm max}$ of
the longest factorable integer can be related to the maximum
acceptable runs of quantum period finding $f_{\rm max}$ and the
smallest accurately implementable controlled rotation gate
$\pi/2^{d_{\rm max}}$ via $L_{\rm max}\sim 4^{d_{\rm
max}-1}\log_{2}f_{\rm max}$. Integers thousands of digits in
length can be factored provided controlled $\pi/64$ rotations can
be implemented with rotation angle accurate to $\pi/256$.
Sufficiently accurate fault-tolerant constructions of such
controlled rotation gates have been described.

\bibliography{References}

\widetext
\end{document}